\documentclass[useAMS,usenatbib]{mn2e}

\usepackage{graphicx}

\title[Breaking the Ice: Planetesimal Formation at the Snowline]{Breaking the Ice: Planetesimal Formation at the Snowline}
\author[G. Aumatell and G. Wurm]{Guillem Aumatell$^{1}$\thanks{E-mail:
guillem.aumatell@uni-due.de} and Gerhard Wurm$^{1}$\\
$^{1}$Fakult\"at f\"ur Physik, Universit\"at Duisburg-Essen, Lotharstr. 1, 47057 Duisburg, Germany}

\begin{document}

\date{Accepted date. Received date; in original form date}

\pagerange{\pageref{firstpage}--\pageref{lastpage}} \pubyear{2002}

\maketitle

\label{firstpage}

\begin{abstract}

Recently \citet[]{Saito2011} proposed that large ice aggregates which drift inwards in protoplanetary disks break up during sublimation, ejecting embedded silicate particles. This would lead to a concentration of small solid particles close to the snowline. In view of this model we carried out laboratory experiments where we observed freely levitating ice aggregates sublimating. We find that frequent break up is indeed very common. Scaled to a 10 cm aggregate about $2 \times 10^4$ small silicate aggregates might result. This supports the idea that sublimation of drifting ice aggregates might locally increase the density of small dust (silicate) particles which might more easily be swept up by larger dust aggregates or trigger gravitational instability. Either way this might locally boost the formation of planetesimals at the snowline.
\end{abstract}


\section{Introduction}

In protoplanetary disks small particles collide, stick together and form larger aggregates \citep[e.g.,][]{Blum2008}. This is usually regarded as a first phase of planet formation. One process with potentially wide impact on planet formation is radial transport of material. In a pressure gradient supported disk the disk rotates somewhat slower than Keplerian. As solids are not supported by this pressure gradient they feel a residual gravity and drift inwards \citep[]{Weidenschilling1977}. In typical disks meter-size bodies drift the fastest moving about 1 AU in 100 y.  Usually, this fast radial drift is regarded as hurdle on the way to planets, as it might lead to the loss of solids to the star, eventually. However, infall of mass from further out also provides means for a local concentration of solids further inward. Two mechanisms of this kind, related to each other, were recently proposed by \citet{Saito2011} and \citet{Sirono2011}. 

The basic idea proposed is that large aggregates drift inwards from the colder regions of the disk where the dominant mass fraction of solids is ice. They sublimate and during this phase fall apart in several pieces. Only the silicate cores embedded within the aggregates survive and due to the fragmentation process a larger number of small particles or aggregates gather at or close to the snowline. As the fragments are smaller in size compared to their progenitor, the drift velocity is much smaller and the particles concentrate around the snowline. As consequence this might boost the formation of planetesimals. The density might rise to a level that gravitational instability leads to the formation of planetesimals \citep{Wakita2008}. Also in a turbulent disk an enhanced dust to gas ratio might be beneficial for planetesimal formation \citep[]{Johansen2009}. Last not least growth efficiencies in the hit-and-stick scenario of planetesimal formation increase as well if larger bodies have a reservoir of small particles to collect (Blum and Wurm 2008). Whatever the details of further evolution, the breakup of large ice aggregates into smaller units while drifting inwards and the resulting concentration of small grains might be an important step in planetesimal and planet formation close to the snowline. Therefore, the models proposed by \citet{Saito2011} 
and \citet{Sirono2011} offer interesting perspectives. The models have some distinctive features.

The model by \citet{Sirono2011} considers sublimation under saturated conditions including sintering. In
a numerical simulation the author shows how water is redistributed in an ice aggregate to the largest grain
within this aggregate, eventually. During this process connections are broken and intermediately the aggregate is split into several smaller aggregates which eventually all sublimate completely.  
This model gets support from laboratory work where sublimating / sintering snow layers under saturated conditions are observed collapsing under gravity \citep{Flin2003}. 

The model by \citet{Saito2011} is similar but considers only sublimation of ice aggregates (no sintering) 
in a water vapor free environment. Here, the basic setup is that silicate cores are 
mantled by ice and these core-mantle grains form
larger aggregates. Reaching the snowline the ice sublimates and all cores are assumed to be ejected as individual grains. This model is not supported by numerical calculations as the model by \citet{Sirono2011}.
The fragmentation of the ice aggregates itself is not discussed but a basic assumption.
In view of this we carried out laboratory experiments for free particles suspended against gravity which very closely resemble the model by \citet{Saito2011} to study how many fragments would
result from sublimation. 

The experiments are based on a levitation technique which was only recently developed. \citet{Kelling2009} showed that porous dust particles levitate over a hot surface at low (mbar) pressure. This is due to thermal creep in a temperature gradient which transports gas  downwards through the pores of the particles. This flow leads to a slight pressure increase below the particle which is sufficient to compensate its gravity. The same effect is used here but at lower temperatures. Details are given below. With this technique it is possible to observe free, levitating ice aggregates while they sublimate and break up which gives a direct measure of the number of times a given aggregate breaks up before it is completely sublimated. The aggregates in our experiments move is a quasi 2d way where gravity and compression due to gravity are only minor factors. 

\section[]{Laboratory Experiments}

\begin{figure}
 \centering
 \includegraphics[width=.47\textwidth, bb= 0 -1 408 428, clip=]{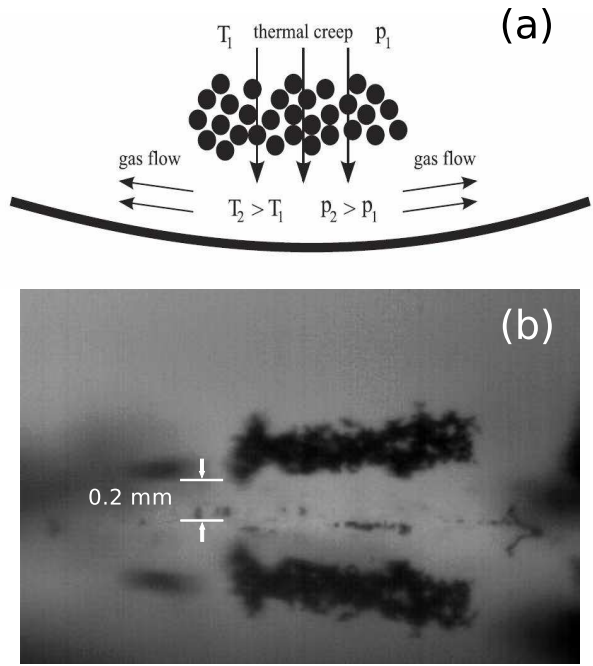}
 \caption{a) Levitation induced by thermal creep (Knudsen compressor). The air flows through the pores from the cold to the warm side building up a pressure (Kelling and Wurm 2009). b) Side view of a levitated ice aggregate on a Peltier element and its reflection.}
 \label{Knudsen}
\end{figure}

We levitate ice aggregates at low pressure (mbar) which allows a sublimation of freely moving 
particles where the breakups of particle parts get directly observable and the number of breakups per time can be counted. The levitation of the particles is based on an overpressure between aggregates and underlying surface \citep{Kelling2009}. This overpressure is generated by thermal creep through the pores of the particle. The thermal creep is induced by a temperature gradient and this temperature gradient is externally generated by cooling the top of the experiment chamber with liquid nitrogen. The pressure increase by thermal creep is also known as Knudsen Compressor \citep{Knudsen1909}. The compression depends on the ambient pressure and allows a particle lift in the mbar pressure range. A particle will levitate at a height where the gas flow to the sides decreases the pressure difference to exactly the level that the weight of the particle is balanced. (see Fig. \ref{Knudsen})

\begin{figure}
 \centering
 \includegraphics[width=.47\textwidth, bb= 0 -1 730 506, clip=]{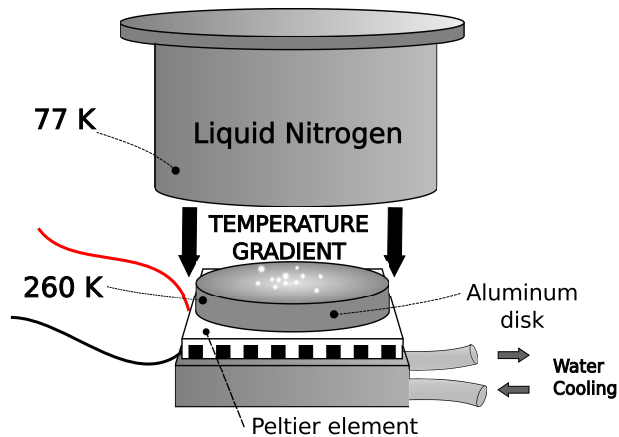}
 \caption{Sketch of the setup used for ice aggregate levitation.  Experiments are carried out at a 
 pressure of few mbar.}
 \label{Setup}
\end{figure}

The actual setup consists of a Peltier element (5cm x 5cm), on which an  aluminum plate is placed. The plate is slightly  concave to confine the aggregates to a limited area. Typical measured temperatures on the aluminum plate surface are 260 K. 
On top of the setup a reservoir of liquid nitrogen at 77 K is placed. A sketch of the setup can be seen in Fig. \ref{Setup}.

\begin{figure}
 \centering
 \includegraphics[width=.44\textwidth, bb= 14 14 665 665, clip=]{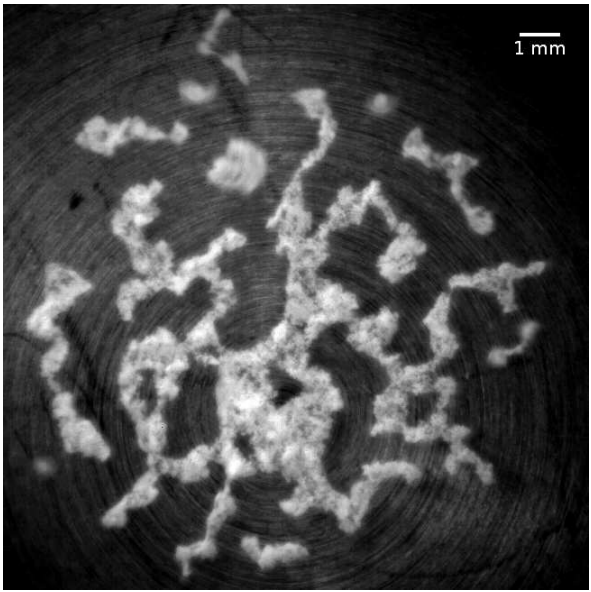}
 \caption{Top view of an ensemble of ice aggregates. }
 \label{Oben}
\end{figure}

\begin{figure}
 \centering
 \includegraphics[width=.45\textwidth, bb= 14 14 627 631, clip=]{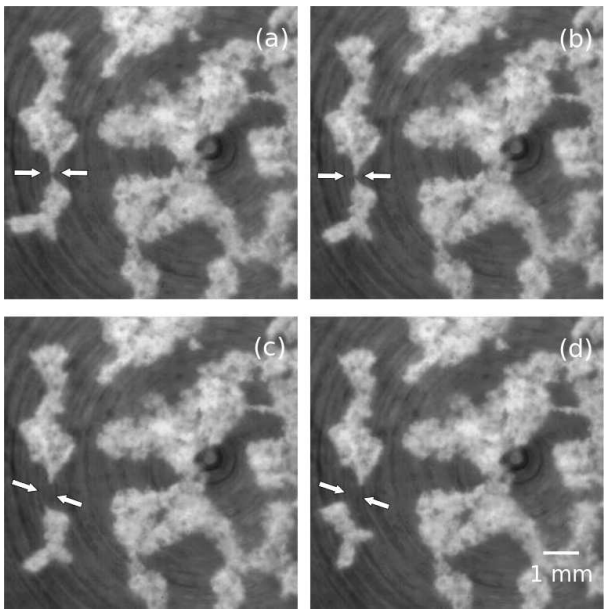}
 \caption{Example of an observed aggregate breakup. In this sequence it can be observed in (a) how the lower part of the left aggregate is still connected to the upper part. In (b) the lower part has displaced slightly to the right but still remains conected. In (c) the aggregate is clearly broken and the lower part separates from the upper one. Finally (d) the two parts will continue their paths separately.}
 \label{Sequence}
\end{figure}

At the low pressure, the ice sublimates at a convinient rate (within a few minutes). As seen in Fig. \ref{Knudsen} (b) the levitating aggregates are typically fractions of a mm in thickness which is small compared to the total area of ice coverage of about 10 mm. This reduces the problem effectively to a 2D problem and breakups can well be observed from a top view of the ice particles (Fig. \ref{Oben}, Fig.\ref{Sequence}). To do so, the nitrogen reservoir has a central clearance (not indicated in Fig. \ref{Setup}).

In the experiments gravity is well balanced by the pressure difference. Only small forces remain which can
lead to an interaction between different particles, i.e. if contacts break aggregate parts are usually drifting apart. Due to the flat and porous structure vertical breakups with a 
succeeding compaction (aggregate parts fall downwards) are expected to be rare though we cannot rule out 
such events. We only consider observable breakups here, where one part of an aggregate clearly separates 
from another part. In each experiment a movie is taken from the sublimating ice particles. On the dark background a separation can easily be seen by eye while scanning frame by frame through the movie afterwards. 
The number of breakups and the times of each breakup are the results of a visual analysis of the movie 
without any further image analysis.

We also count 
breakups where two aggregate parts separate but are still connected at another location, i.e. bend and open at the breakup point. 
However, such events only account for about 10\% of all cases. This is in agreement with the formation
of the aggregates by hit-and-stick growth during preparation, where every individual ice grain only has two neighbors. 

\begin{figure}
 \centering
 \includegraphics[width=0.47\textwidth, bb= 0 0 1022 620, clip=]{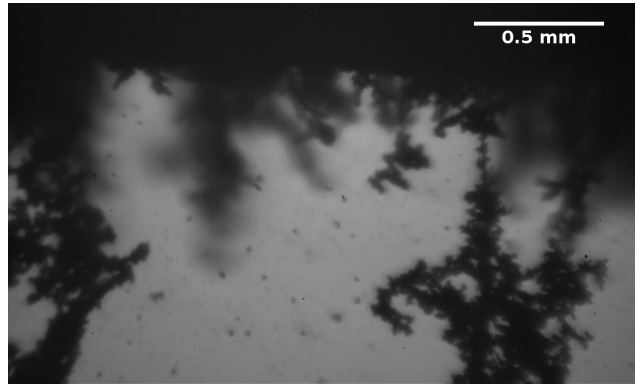}
 \caption{Example from the preparation process. Atmospheric water vapor condenses into ice grains (small particles) around a cold liquid nitrogen pipe (top black area). They collide and stick to aggregates which
 stick to the pipe. }
 \label{Hanging}
\end{figure}

Particle Preparation: Ice particles were prepared under atmospheric conditions (1 bar, laboratory humidity). 
Individual ice grains condensate in the vicinity of the pipe's surface that connects the liquid nitrogen tank and the setup's liquid nitrogen reservoir. These ice grains 
collide, stick together and form aggregates, which can be observed by microscopy. Fig. \ref{Hanging}
shows a snapshot of the preparation phase. We did not specify the morphology or structure of
the individual grains in more detail. The aggregates are taken from the pipe's frozen surface and put into the setup using a small spoon cooled by liquid nitrogen. Also, the transfer of particles into the 
experiment chamber might induce modifications e.g. by direct condensation of water vapor but essentially the ice particles observed in the experiments are grown aggregates.

As aggregates the
particles are an appropriate model for large particles in protoplanetary disks which
are also assumed to be formed by aggregation \citep{Blum2008}.

\begin{figure}
\centering
\includegraphics[width=.45\textwidth, bb= 0 -1 682 405, clip=]{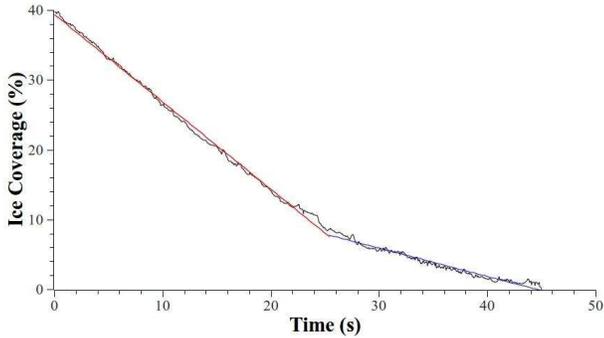}
\caption{Evaporation for a selected experiment. Shown is the fraction of the image covered by ice over time. A change in slope is visible at 25 s. }
 \label{Sublimation}
\end{figure}

An example of the evaporation rate is seen in Fig. \ref{Sublimation} which shows the evolution of the evaporating ice during an individual experiment as fraction of the image covered by ice. The
sublimation can well be described as linear with time but with a pronounced change in slope at about 25 s. Within the scope of this paper we consider either one or two linear approximations for all sublimation curves. We then get constant sublimation rates

\begin{equation}
a=\frac{dA}{dt}= \rm const
\end{equation}
where \textit{a} is $-1.25 \pm 0.01$ $\mathrm{mm^2/s}$ initially ($<$ 25 s  in Fig. \ref{Sublimation}) 
and $-0.41 \pm 0.01$ $\mathrm{mm^{2}/s}$ for the later times in the given example. This is a difference of about a factor 3.
 
\begin{figure}
\centering
\includegraphics[width=.45\textwidth, bb= 0 -1 686 450, clip=]{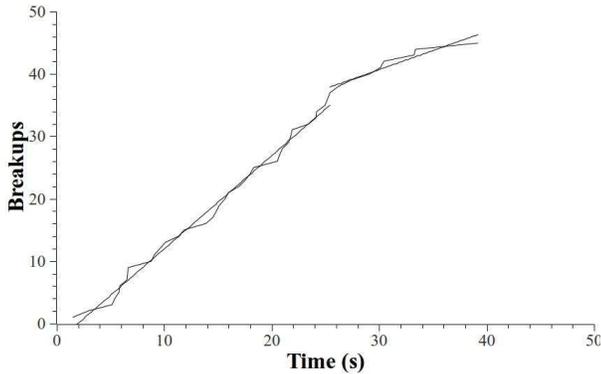}
\caption{Number of breakups over time. Also here a linear dependence can be observed which is split in time in two parts. The split is correlated to the split
in evaporation rate of the same experiment (see Fig. \ref{Sublimation})}
 \label{breakstime}
\end{figure}

For the breakup rate (number of observed breakups per second) we observe similar splits in two parts as for the evaportation rate which are correlated in time.
This is shown in Fig. \ref{breakstime}.

To quantify the total number of breakups for a given size we measure the breakup rate and the total time it takes the aggregate to sublimate from the
initial size.  This time is determined by the sublimation rate \textit{a} and the inital particle size.
As initial particle size we considered the two dimensional radius of gyration of the particle system at the beginning of an experiment defined as
\begin{equation}
r=\sqrt{\frac{\sum_{i}{\left(r_i - r_s\right)}^{2}}{N_p}}
\end{equation}
with $r_i$ being the coordinates of the i-th pixel covered by ice and $r_s$ being the coordinates of the center of mass and $N_p$ is the total number of pixels. 
  
\begin{figure}
 \centering
 \includegraphics[width=.45\textwidth, bb= 0 -1 641 381, clip=]{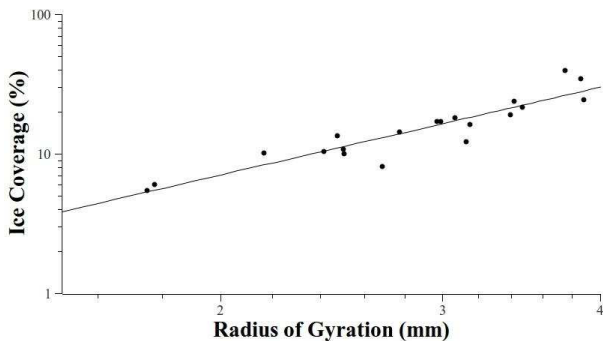}
 \caption{Area covered by ice over the size of an aggregate at the beginning of an experiment. Each data point represents an individual experiment.}
 \label{Icecoverage}
\end{figure}

Fig. \ref{Icecoverage} shows the ice surface coverage $A$ over size $r$. The data follow a power law with power $ d=2.0 \pm 0.2 $.
This is consistent with a porous but otherwise uniform coverage of ice in 2 dimensions independent of size or $d=2$.  Taking $d=2$ the ice covered area can
be approximated by $A = br^2$ with $b= 2.20 \pm 0.26 $. 

In the cases where we have to consider two distinct breakup 
rates we relate the second breakup rate to the size at the time the breakup
rate changes  (Fig. \ref{breakstime}). If a change in sublimation rate is not obvious (mostly for the smaller aggregates) only one average breakup rate and one size are considered. 

\begin{figure}
 \centering
 \includegraphics[width=.45\textwidth, bb= 0 -1 706 413, clip=]{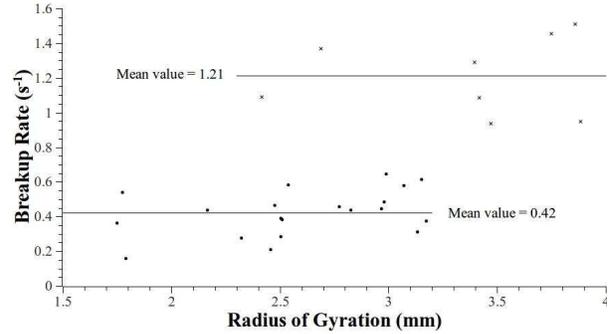}
 \caption{The breakup rate for ice particles with different initial size. Upper and lower values are correlated to the difference in sublimation rate. Each data point corresponds to a different experimental run. Solid 
 lines mark average rates.}
 \label{Breakrate}
\end{figure}

Fig. \ref{Breakrate} shows the breakup rate dependence on size. Upper and lower values are correlated to the difference in sublimation rate. 
The average breakup rate differs by about a factor of 3 which is the same factor as for
the correlated difference in evaporation rate. 

Fig. \ref{Breakrate} suggests that the breakup rate is not dependent on the
aggregate size.  We would like to caution 
though that the
breakup constancy is not a universal feature in detail. 
As described above, aggregates form during the preparation of the ice particles. Individual
grains have a size distribution between several up to a few tens
of microns, though the latter might already be small aggregates 
(Fig. \ref{Hanging}). This size distribution directly translates to a distribution in contact areas or necks 
between neighboring grains. We cannot
spatially resolve these contact areas. For aggregates with a
distribution of neck sizes the breakups proceed continuously in
time with some statistical variations. Besides this the neck size distribution determines the breakup rate in detail. It is the inability to quantify this distribution that the breakup rate 
of \textit{all experiments} is approximated by one average value. 
We note though that average breakup rates
in \textit{individual experiments} with similar initial size might differ by a factor of
3 (see fig. \ref{Breakrate})! More complex breakup rate dependencies on size or even for
a given size
would certainly be possible within the
uncertainties of the data.
However, these simplifications allow us to specify a total
number of breakups for a given size and scale this to much larger aggregates in
protoplanetary disks.

\section[]{Discussion}
The experiments give direct evidence that sublimating ice aggregates 
frequently break up as has been proposed by Saito 
and Sirono (2011) and -- including sintering -- by Sirono (2011). This will generate a number of smaller aggregates which 
eventually reduce to aggregates of embedded silicate (dust) particles. This is an interesting novel idea on the road to 
explain the formation of planetesimals. Our laboratory experiments support this showing 
that breakup is indeed a viable mechanism to increase the density of smaller particles in the vicinity of the snowline. 
We consider the ice in our experiments as
suitable analog to cluster-cluster aggregates in protoplanetary disks. It is likely that some 
initial difference in ice structure and morphology is responsible for the two different 
sublimation rates observed. For each of the two existing sublimation branches we assign an average breakup rate, \textit{n}. 
The total number of breakups, \textit{N}, is therefore $ NT $, where \textit{T} is the total 
sublimation time. The given average surface sublimation rate (on each of the two branches) $ a=dA/dt $ gives a linear total 
time T with aggregate surface, $ T=A/\left | a \right| $. In terms of size this translates to a $ T=br^{2}/\left | a \right| $. 
The total number of breakups then is

\begin{equation}
N=\frac{nbr^{2}}{\left | a \right |}=\chi r^{2}
\label{finalbreaks}
\end{equation}

As $n$ scales with a factor 3 in the experiments correlated to the same scaling factor for $a$ the actual sublimation rate is not important and we combine all empirical factors in a single constant parameter $\chi  \simeq$ 2 $mm^{-2}$in eq. \ref{finalbreaks}.
It had to be expected that the total number of breakups does not depend on time or any time dependent breakup rates or total sublimation rates. So the result that the total number of breakups only depends on the size is consistent with these expectations.

The laboratory situation is effectively a 2d scenario where the 2d projection covered by ice is proportional to the square of the size. Ice aggregates in 3d have the same projected area if they have a fractal dimension of 2. This corresponds to aggregates which grow via cluster-cluster aggregation (CCAs) where aggregates of the same size hit and rigidly stick to each other  at the first contact point. This growth process initially occurs in protoplanetary disks  (Blum and Wurm 2008). For fractal dimension 2 the projected surface essentially also scales with dimension 2. That means
that the size of protoplanetary ice aggregates correlates well with the
size of aggregates in our experiments. Therefore our results can directly be applied to large aggregates in protoplanetary disks and eq. \ref{finalbreaks} gives the number of breakups for a CCA cluster of given 3d size $r$.

Assuming the largest ice aggregate in the form of a CCA would be about 10 cm in a protoplanetary disk then the aggregate would fragment into $\sim 2 \times 10^{4}$  pieces, a 1 cm CCA would fragment into $\sim$ 200 pieces, releasing a large number of small dust aggregates, eventually. More compact ice aggregates will certainly fragment as well but the number of total breakups might vary. 

The laboratory experiments are closely resembling drifting particles in protoplanetary disks. Certainly, in detail the total number of breakups might differ from the laboratory value. Nevertheless, 
eq. \ref{finalbreaks} might be a suitable approximation from the experimental side. 
Clearly not all dust particles embedded in an ice aggregate will leave as single particles as
the shrinking ice aggregate only forms a limited number of smaller ice aggregates.
We assume (not the result of this study) that every ice fragment with silicate cores will 
eventually be reduced to an aggregate of its silicate cores with the same morphology, i.e. a CCA ice aggregate results in a CCA silicate aggregate.

The principle idea behind the proposals by \citet{Saito2011} 
and \citet{Sirono2011} is that the released silicate particles drift slower.
Actually, for CCAs the overall aggregate size does not change the gas-grain-coupling time 
(Blum and Wurm 2008).
As the gas-grain coupling time determines the drift velocity such CCAs would always drift with 
the same speed. This is due to the fact that the gas-grain coupling time is proportional to $M/A$, where 
$M$ is the aggregate mass and $A$ is the aggregate cross
section or surface. For CCAs consisting of the same individual grains $M/A$ is constant.
However, the individual grain size of the cores is - per definition - smaller than the size of the 
individual ice grains. Also the density of silicates is about twice or three times as high as the density 
of water ice. 
This decreases the $M/A$ ratio by a certain factor.
Therefore, these final dust aggregates couple better to the gas of the disk and their radial drift velocity is indeed reduced. 

For compact aggregates (not subject to this study) breakup might also occur. 
If final silicate aggregates are compact the mass scales with $M \sim r^3$ and $M/A$
depends linear on the overall size of the aggregates. Any smaller fragment then drifts slower than
the original large ice aggregate. 

Therefore, while the details depend on the initial ice aggregate morphology and the details of 
the break up including silicate cores the break up of ice aggregates due to sublimation at the snowline 
will occur and it almost inevitably leads to a density enhancement of silicates.

\section{Conclusion}

Saito and Sirono (2011) and Sirono (2011) proposed that ice aggregates drifting inwards and sublimating at the snow line leave behind a large number of smaller silicate or dust grains with much smaller radial drift velocity. This leads to a dust density enhancement close to the snowline and subsequently an increased likelihood for planetesimal formation. Planetesimal formation might be due to more frequent sticking collisions and growth or due to gravitational instabilities. The proposal by Saito and Sirono (2011) was based on the mere assumption that ice aggregates break up upon sublimation and release silicate particles. We verified these ideas of ice aggregate breakup upon sublimation in laboratory experiments. Indeed we find that ice aggregates break up in a large number of smaller aggregates. Scaled to protoplanetary disks a 10cm CCA aggregate would fragment into $\sim 2\times 10^{4}$ pieces. We only observed the fragmentation of pure ice aggregates. We currently assume that all silicate particles that would be embedded in one ice fragment would end up as one aggregated particle after total sublimation, i.e. a 10cm CCA ice aggregate would result in $\sim 2\times 10^{4}$ silicate aggregates. If the presence of silicate cores would
influence the sublimation process or what the morphology of the silicate aggregates looks like is subject to future studies. However, the observed breakup into smaller fragments supports the idea that sublimation of ice aggregates close to the snowline might promote the local formation of planetesimals by providing a reservoir of silicate particles that drift slower.

\section*{Acknowledgments}

G. Aumatell is supported by the European Community’s Seventh Framework Programme (FP7/2007-2013 under grant agreement No. 238258). We appreciate the valuable comments from the reviewer.

\bibliographystyle{mn2e}
\bibliography{References}

\label{lastpage}

\end{document}